# SMALL FOOTPRINT NANO-MECHANICAL PLASMONIC PHASE MODULATORS

*V.A. Aksyuk[1], B.S. Dennis[1,2], M.I. Haftel[3], D.A. Czaplewski[4], D. Lopez[4], and G. Blumberg[2]*
[1]Center for Nanoscale Science and Technology, National Institute of Standards and Technology, Gaithersburg, MD, USA
[2]Rutgers, the State University of New Jersey, Dept. of Astronomy and Physics, Piscataway, NJ, USA
[3]University of Colorado, Dept. of Physics, Colorado Springs, CO, USA
[4]Argonne National Laboratory, Center for Nanoscale Materials, Argonne, IL, USA

## ABSTRACT

The authors' recent *Nature Photonics* article titled "Compact Nano-Mechanical Plasmonic Phase Modulators" [1] is reviewed which reports a new phase modulation principle with experimental demonstration of a 23 μm long non-resonant modulator having 1.5 π rad range with 1.7 dB excess loss at 780 nm. Analysis showed that by decreasing all dimensions, a *low loss,* ultra-compact π rad phase modulator is possible. Application of this type of nano-mechanical modulator in a miniature 2 x 2 switch is suggested and an optical design numerically validated. The footprint of the switch is 0.5 μm x 2.5 μm.

## INTRODUCTION

Device demonstrations based on collective optical frequency electronic oscillations localized to single metal-dielectric interfaces, or *surface* plasmons [2], are numerous [3]. Increasingly, plasmonic devices use *two* metal-insulator interfaces, separated by a narrow gap across the insulator layer, that confine *gap* plasmons (GP) in metal-insulator-metal (MIM) waveguides [4] to study extreme confinement and the dependence of the effective refractive index on gap [5]. With increasing confinement, local fields are enhanced, phase velocity is decreased, while a larger fraction of the energy is transferred from the insulator into the surrounding metals. Varying the gap provides a way to change the effective refractive index of an MIM plasmonic device to electrically control the GP, allowing a control mechanism potentially useful in photonic switching fabrics and reconfigurable plasmonic optics.

Photonic switches reconfigure networks and communication channels and often rely on phase modulation elements. Switching is typically slower than data modulation, with emphasis put on low power usage and optical losses, small size, and broad optical bandwidth. Smaller, lower power switching fabrics with 1 μs to 10 ns switching times are desired to enable dynamic reconfiguration in photonic architectures.

Many modulation principles have been used to demonstrate a diverse range of photonic phase modulators such as thermo-optical devices [6], very fast slot plasmon electro-optical devices [7], and electro-mechanical devices.

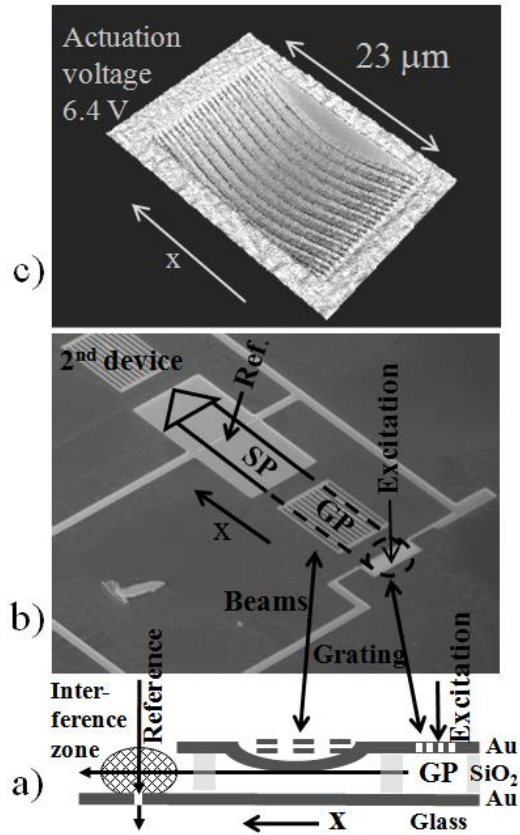

*Figure 1:* a) Schematic of the Gap Plasmon Phase Modulator (GPPM). Laser light grating-couples into the waveguide, propagates to the left through the transparent support pillars and under the beams, interferes with the reference beam, and out-couples below where it is collected by a microscope. b) SEM of the device. c) Interferogram of a device with beams electrostatically actuated with 6.4 V (beam displacement exaggerated for clarity). GP is gap plasmon and SP is surface plasmon.

Optically resonant electro-optical and electro-mechanical devices trade phase-modulation strength for reduced wavelength range. Semiconductor and plasmonic devices based on carrier concentration change tend to have large absorption modulation that results in high excess loss for phase modulation.

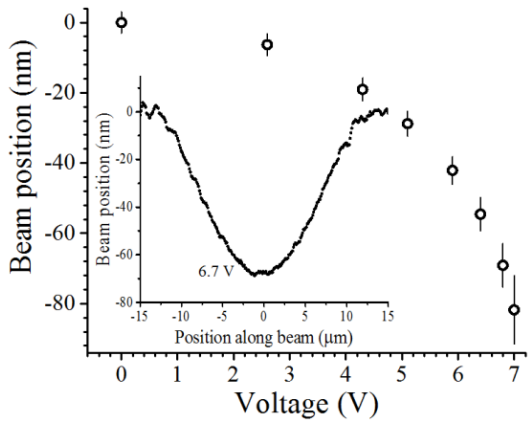

*Figure 2: Nano-mechanical **b**eam displacement as a function of applied voltage. Inset shows the profile of one beam actuated with 6.7 V. Error bars are single standard deviation uncertainties, see [1].*

In their new *Nature Photonics* article [1] the authors have proposed a nano-mechanical solution to tune the refractive index of an MIM waveguide that exploits the strong dependence of the phase velocity of confined GPs on dynamically variable gap size. A micro-fabricated 23 μm long non-resonant **Gap Plasmon Phase Modulator (GPPM)** allowed electrical control of the GP phase. Phase modulation of 1.5 $\pi$ rad with 1.7 dB excess loss was demonstrated interferometrically. A computational model was presented demonstrating the realization of an ultra-compact footprint $\pi$ rad GPPM by decreasing the device's length, width, and initial MIM gap. Surprisingly, the extra loss expected for GPs confined in a decreasing gap was offset by the increasing phase-modulation strength gained from the same deceasing gap.

A switching application of the GPPM is proposed and numerically modeled. The Mach-Zehnder optical switch has an extremely small footprint with modest optical loss enabled by the strong phase modulation of GPs in a mechanically actuated 17 nm air gap. Frequency-domain finite-element modeling at 780 nm showed that the insertion loss is 8.5 dB, the extinction ratio is > 25 dB, and crosstalk for all ports is > 24 dB.

## GPPM FABRICATION

A Au/SiO2/Au stack with sputtered Au and plasma enhanced chemical vapor deposition $SiO_2$ layers, all three (220 ± 5) nm thick, was deposited onto nominal 500 μm thick glass with an ≈ 10 nm Cr adhesion layer between the substrate and bottom Au layer and an ≈ 2 nm thick Ti adhesion layer on both sides of the $SiO_2$. Device patterns were written with e-beam lithography. After resist development, device components were Ar ion milled into the top Au layer. The beams were released by wet etching of the underlying $SiO_2$ in buffered oxide etch with subsequent $CO_2$ critical-point drying. The $SiO_2$ was completely removed everywhere below the lithographic patterns leaving a lateral undercut of ≈ 2.5 μm. After release, the $SiO_2$ pillars supporting the beams at their ends were ≈ 3 μm wide in the direction of GP propagation. The out-coupler slit was ≈ 150 nm wide by ≈ 20 μm long and was cut with a focused ion beam. The in-coupler grating was composed of strips ≈ 18 μm long and ≈ 400 nm wide with period either ≈ 720 nm or ≈ 760 nm.

## GPPM EXPERIMENTAL

Figure 1a shows a schematic of the GPPM. The electrostatically tunable gold-air-gold waveguide has a device dependent initial air gap around 280 nm. The top gold film is patterned into eleven suspended deformable metal beams, each (23.0 ± 0.5) μm in length and (1.50 ± 0.07) μm wide supported at both ends by $SiO_2$ pillars. A GP, launched via grating coupling with a focused free-space excitation laser, propagates underneath and along the beams. A focused reference beam, split from the excitation laser and incident at 13.2°, interferes with the plasmon at the out-coupler slit.

As shown in Figures 1c and 2, when a voltage is applied, the electrostatic force deforms the bridges down into an approximately parabolic shape, narrowing the MIM gap at the beam center by about 80 nm as the voltage increases up to a maximum of 7 V and phase-retards the GP. To measure the GPPM optical performance, an out-coupler slit is used to sample the modulated plasmon using a microscope from below. The tilted-reference laser allows phase-sensitive imaging of the modulated plasmon. Using a Mach-Zehnder type interferometer both the GP phase retardation and optical loss are measured as a function of gap by electrostatically controlling the GPPM beam displacements. Optical images of the out-coupled light are collected with and without the reference laser at different applied DC voltages.

Figure 3 shows the change in phase induced as a function of waveguide height by applying 0.0 V to 7.0 V to the GPPM. This device demonstrated the largest phase change. The waveguide height is measured at the beam center. The transmission, or excess optical power loss, caused by the narrowed gap under the actuated beams can be seen in Figure 3 inset, which plots the integrated areas of Gaussian intensity fits normalized by that of the unactuated device. A phase shift exceeding 1.5 $\pi$ rad was achieved, while the corresponding excess loss is near 30 % (1.7 dB) when the gap is tuned by approximately 30 %, from 270 nm to 190 nm. The uncertainties indicated throughout the paper are +/- one standard deviation, with detailed uncertainty analysis presented in [1].

## GPPM SCALING ANALYTICS

MIM waveguides support guided modes for frequencies below the surface plasmon resonance and for gaps near the single nanometer range: local classical theory begins to break down below that. Also, the effective refractive index of these waveguides dramatically increases and the GP wavelength decreases in small gaps. Furthermore, the phase modulation strength in this geometry increases approximately inversely with the square of the gap, making it exquisitely sensitive to nanoscale motion for sensing and desirable for on-chip optical actuation in applications where strong yet broadband optomechanical coupling is required. Decreasing the initial gap increases optical propagation losses, as more of the optical power resides in the metal. As seen in [1], decreasing the beam length and gap by an appropriate amount (and hence the optical travel distance) allows the loss through an unactuated device (insertion loss) to remain constant, e.g. at 1/e power (4.3 dB). The relationship is that for each beam length there is an initial gap where the length scales as the initial gap to the 0.8 power. The impact of this is that if the GPPM dimensions are scaled down as described, the phase modulation range will be maintained, without incurring a loss penalty. The result of simultaneously reducing both the length and initial gap, by more than an order of magnitude than the demonstrated GPPM, is shown in the plots of the calculated change in phase and transmission vs. waveguide height (Figure 4). Interestingly, the phase modulation range stays constant with miniaturization for a given optical loss. For an initial gap much smaller than the SP evanescent decay distance, universal scaling emerges between the phase shift and the excess loss such that they are linearly related regardless of the initial gap, e.g. as the gap is decreased to 72 % of the initial gap, π rad phase modulation is maintained with excess loss of 0.8 dB independent of the device scale [1].

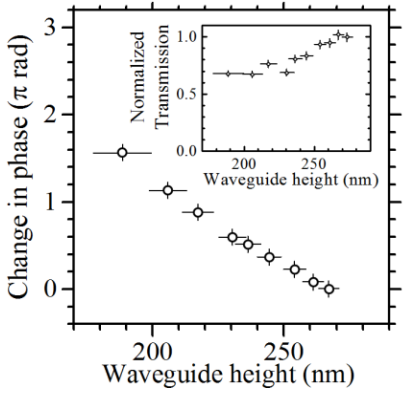

*Figure 3: Change in phase plotted vs. waveguide gap for the GPPM which gave the maximum phase shift (≈ 1.5 π rad). Inset: Normalized transmission vs. gap ( same device). Error bars are single standard deviation uncertainties, see [1].*

## 2 x 2 MACH-ZEHNDER SWITCH

One application of the GPPM, with potential use as a photonic switching fabric element, is a 2 x 2 Mach Zehnder switch (Figure 5). The device has two input ports that connect to two arms via a 50/50 (3 dB) coupler, where at least one arm is nanomechanically phase modulated. The arms connect to two output ports through another 3 dB coupler. One common way to incorporate the plasmonic phase modulator into a photonic circuit would be to use dielectric waveguide technology, such as SOI (silicon on insulator). Coupling dielectric waveguides with low-loss to nanoscale plasmonic gaps has been reported [8] but since these systems cannot be further miniaturized, the full potential for downscaling inherent in plasmonics is lost.

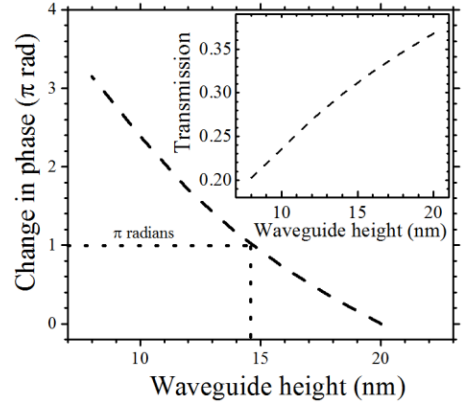

*Figure 4: Calculated change in phase vs. waveguide height for an initial 20 nm gap. The beam length was chosen to give a 1/e (-4.3 dB) initial insertion loss. An actuation depth of ≈ 72 % results in a phase shift of π rad (dotted line) and avoids pull-in. Inset: Calculated transmission vs. gap using the same beam length.*

It is therefore necessary to incorporate such nanoscale components directly for gap plasmons and to tightly integrate them into completely plasmonic switches. We propose and numerically demonstrate a design for such an integrated 2 x 2 switch. The switch's performance is calculated using a full-vector three-dimensional (3D) frequency-domain finite-element-model (FEM). These 0.5 µm$^2$ (0.25 µm x 2 µm) devices achieve π rad phase modulation and have a smaller footprint than the GPPM in [1], even including the lateral air gaps on the sides. Surprisingly, a modest loss of ≈ 8.5 dB with an extinction ratio greater than 25 dB is calculated.

The switch model uses an MIM gold/air/gold stack (Figure 5) and can be fabricated and electrostatically actuated similarly to the GPPM [1]. Different from the GPPM, an additional structural layer, that is not shown, caps the top gold, and is structured such that only one modulator arm is moveable, while the reference arm and couplers are held fixed. As in [1] a parabolic profile is used to appoximate the deformation shape of the nanomechanically movable arm. The color scale in Figure 5 indicates the vertical deformation depth. In a complete switch design, both the electrostatic and mechanical actuation would need to be analyzed fully, and a step by step fabrication process sequence would need to be developed, but the goal of this work is to model the optical performance of such switches, where a parabolic approximation of deformation is appropriate.

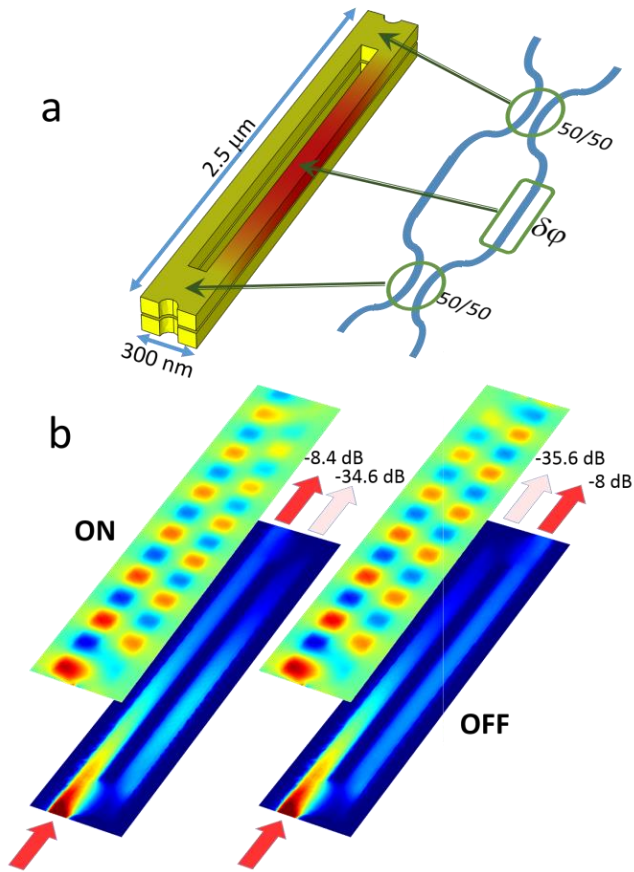

*Figure 5: Nanomechanical gap plasmon 2 x 2 optical switch. a) Perspective view and schematic. Modulator arm actuation indicated by color scale. Schematic shows the light paths between input and output ports through 3 dB couplers and phase modulation and reference arms. b) Instantaneous values of the horizontal magnetic field component normal to the direction of GP propagation (upper), and local power flow in the direction of propagation (lower), in the mid-plane of the gap in the switch ON and OFF states under port 1 excitation. Transmission to each of the output ports is marked for each case.*

Uncertainties estimated at ≈ 1 dB in the calculated loss and crosstalk are based on FEM sensitivity to mesh density. These numerical errors are small compared to real experimental factors such as the surface finish of the metal, dielectric constant variability with deposition parameters, and geometrical effects due to imperfect structure shapes.

## CONCLUSION

The authors have experimentally demonstrated strong optomechanical transduction with low optical losses in electrostatically actuated nanoscale gap MIM plasmonic modulators. The 23 μm long GPPMs, with an average optomechanical modulation strength of 52 mrad/nm at 780 nm, achieved a maximum of 5 rad of phase modulation with low insertion and excess losses. An analytical model in good agreement with measurements shows miniaturization to a sub 1 μm$^2$ footprint, without degradation in optical performance and with an increase in speed and decrease in actuation voltage, is possible.

This new concept enables a new class of on chip optical switching and optical circuit reconfiguration functionality. A 2 x 2 nanomechanical plasmonic switch based on the GPPM is both proposed and modeled. The ultra-compact low power switch has a 0.5 μm$^2$ footprint with a greater than 25 dB extinction ratio and a modest 8.5 dB loss.

## ACKNOWLEDGEMENTS

The authors acknowledge support from the Measurement Science and Engineering Research Grant Program of the National Institute of Standards and Technology under award numbers 70NANB14H259 and 70NANB14H030 and Air Force Office of Scientific Research under Grant No. FA9550-09-1-0698. Computational support of the Department of Defense High Performance Computation Modernization project is acknowledged. This work was performed, in part, at the Center for Nanoscale Materials, a U.S. Department of Energy, Office of Science, Office of Basic Energy Sciences User Facility under Contract No. DE-AC02-06CH11357.

## CONTACT

V.A. Aksyuk, tel: +1-301-975-2867; vladimir.aksyuk@nist.gov